\documentclass[11pt,twoside]{article}
\usepackage{asp2010}

\resetcounters

\begin{document}

\title{Stellar Evolution with Pulsation-Driven Mass Loss:
The Case of LMC Cepheids}
\author{Hilding R. Neilson$^1$, Norbert Langer$^1$, and Matteo Cantiello$^1$
\affil{$^1$Argelander-Institut f\"{u}r Astronomie, Auf dem H\"{u}gel 71, D-53121 Bonn, Germany}
}

\begin{abstract}
The Cepheid mass discrepancy, the difference between mass estimates from stellar pulsation and stellar evolution models, is a long standing challenge for the understanding of stellar astrophysics. We discuss the current state of the mass discrepancy and test the role of pulsation-driven mass loss in state-of-the-art stellar evolution calculations of Large Magellanic Cepheids in resolving it.  We find that Cepheid mass loss is a significant contributor to the mass discrepancy, but it is not clear if the metallicity dependence of Cepheid mass loss is consistent with the measured metallicity dependence of the mass discrepancy.
\end{abstract}

\section{Introduction}
Classical Cepheids are ideal laboratories for understanding stellar astrophysics and evolution.  Observations of Cepheid pulsation probes the interior structures of these stars constraining stellar evolution theory, while their luminosities $\log L/L_\odot > 3$ allows us to observe Cepheids in distant galaxies with differing metallicities.  Furthermore, the tight correlation between the pulsation period and stellar luminosity also makes them powerful standard candles.  
However, for all their importance for stellar astrophysics and cosmology, the masses of Cepheids are still not understood.  

Cepheid masses are determined in three ways: from stellar evolution models, stellar pulsation models, and from measurements of binary systems with a Cepheid component.  For example, \cite{Piet2010} determined the mass of the only known fundamental-mode Cepheid to be in an eclipsing binary.  Other dynamical masses have been determined for Cepheids in other types of binary systems, but not to the same precision \citep[e.g.][]{Evans2008, Evans2009}.  Masses from stellar evolution models are determined by fitting evolutionary tracks to the observed effective temperature and luminosity, while masses from stellar pulsation models are determined by fitting pulsation properties.  The resulting paradox is that these latter two methods predict different masses, this is historical problem is called the Cepheid mass discrepancy \citep{Cox1980}.

Currently, the Cepheid mass discrepancy has been measured for Galactic, and Large and Small Magellanic Cloud Cepheids.  \cite{Keller2006} and \cite{Keller2008} determined that the mass discrepancy ranges from about $15$-$25\%$, where the mass discrepancy is defined as the relative difference between evolutionary and pulsation predictions.  Furthermore, they find that the mass discrepancy increases decreasing metallicity.  It should be noted that stellar evolution masses in these works are computed assuming no extra mixing processes such as rotational mixing or convective core overshooting.  

\cite{Bono2006} reviewed possible solutions to the Cepheid mass discrepancy, where the two most likely candidates are mass loss during the Cepheid stage of evolution and convective core overshooting in the main sequence progenitors of Cepheids.  Convective core overshooting is defined as the distance above the stellar core that convective eddies penetrate and mix over an evolutionary timescale.  This distance is typically defined as $\Lambda = \alpha_c H_P$, where $H_P$ is the pressure scale height and $\alpha_c$ is a free parameter.  Overshooting mixes additional hydrogen into the core of a main sequence progenitor, producing a more massive helium core during the Cepheid stage of evolution.  Therefore, by including overshooting one can fit observed luminosity of a Cepheid with a stellar evolution track assuming smaller masses than an evolution model with no overshooting, and thus brings evolution calculations into agreement with pulsation models.  However, \cite{Keller2008} found that a mass discrepancy of $20\%$ requires a value of $\alpha_c = 0.8$.  This value of $\alpha_c$ is inconsistent with observations of other stars such as eclipsing binaries \citep{Sandberg2010}, $\beta$ Cephei stars \citep{Lovekin2010}, and massive B-type stars \citep{Brott2011}.

Mass loss is another possible solution; \cite{Neilson2008,Neilson2009} derived an analytic theory describing pulsation-driven mass loss in Classical Cepheids, and found that mass-loss rates could be enhanced by more than three orders-of-magnitude relative to a radiation driven wind. Observational evidence for pulsation-driven mass loss is growing; \cite{Neilson2009a,Neilson2010} measured mass-loss rates from infrared excess for LMC Cepheids in the OGLE-II/III surveys combined with the SAGE survey \citep{Ngeow2008,Ngeow2009}. \cite{Marengo2010} and \cite{Barmby2011} determined mass-loss rates, $\dot{M} = 10^{-8}$ to $10^{-7}~M_\odot~$yr$^{-1}$, for Galactic Cepheids from infrared observations, while \cite{Matthews2011} measured a mass-loss rate of $10^{-6}~M_\odot~$yr$^{-1}$ for $\delta$ Cephei, based on 21 cm observations.  This evidence is pointing towards enhanced mass loss in Cepheids and is thus worth testing if Cepheid mass loss resolves the mass discrepancy.

In this work, we compute stellar evolution models including the \cite{Neilson2008} Cepheid mass loss prescription with standard LMC metallicity and assuming moderate convective core overshooting from \cite{Brott2011}, $\alpha_c = 0.335$, to compare with the results from \cite{Keller2006}.  The models are computed through blue loop evolution and we determine the change of mass due to Cepheid mass loss.  We test if this amount of overshooting and mass loss is consistent with the measured mass discrepancy.  Furthermore, we compare the results from LMC Cepheids with the results for Galactic metallicity \citep{Neilson2011}.

\section{Method}
We compute stellar evolution models using the \cite{Yoon2005} version of the state-of-the-art \cite{Heger2000} stellar evolution code.  In this code, we include a prescription for Cepheid mass loss \citep{Neilson2008} that computes pulsation-driven mass loss when the effective temperature and luminosity of the stellar model falls within the Cepheid instability strip as defined by \cite{Bono2000}.  The mass-loss theory predicts the mass-loss rate as a function of luminosity, mass, radius, pulsation period and pulsation amplitudes. 
 The pulsation period is determined from the period-mass-radius relation \citep{Gieren1989}, and the pulsation amplitudes for luminosity and radius are from \cite{Klagyivik2009}.  Details can be found in \cite{Neilson2011}.  We note that these relations are derived for Galactic Cepheids, hence Galactic metallicity, and thus ignores any implicit metallicity dependence of the pulsation period and amplitudes.
 
 \section{Results}
 
\begin{figure}[t]
\begin{center}
\includegraphics[width=0.5\textwidth]{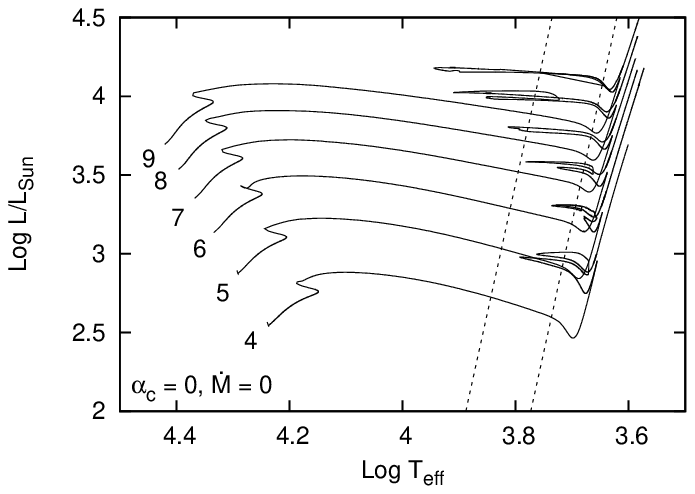}\includegraphics[width=0.5\textwidth]{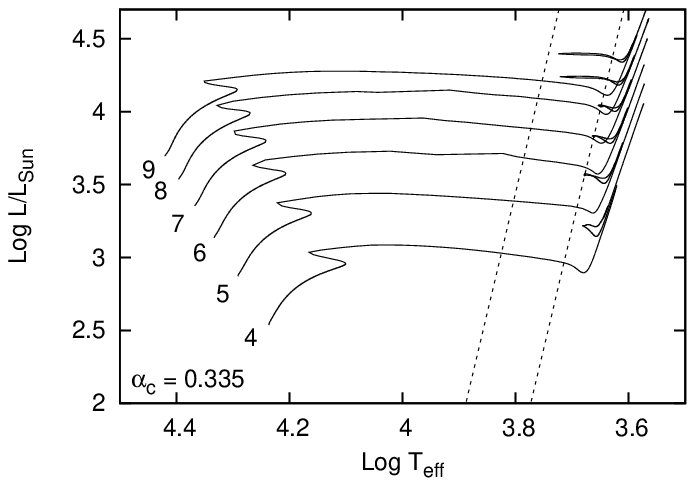}
\end{center}
\caption{Stellar evolution tracks for models with Galactic metallicity, $Z=0.02$ for two scenarios.  The first (left) is for models assuming no mass loss and $\alpha_c = 0$,  and the second (right) is for stellar evolution models including the \cite{Neilson2008} Cepheid mass loss prescription and $\alpha_c = 0.335$. The dashed lines outline the boundaries of the Cepheid instability strip \citep{Bono2000}.}\label{f1}
\end{figure}

 In \cite{Neilson2011}, we compute stellar evolution models with Galactic metallicity, $Z=0.02$, and masses $M=4$ to $9~M_\odot$ in steps of $1~M_\odot$ for four scenarios.  The first scenario is no mass loss and no convective core overshooting, while remaining scenarios include pulsation-driven mass loss and convective core overshooting with values of $\alpha_c = 0, 0.1$, and $0.335$.  We show the stellar evolution tracks for models with no mass loss and zero convective core overshooting include along with models with pulsation-driven mass loss included and $\alpha_c=0.335$ in Fig.~\ref{f1}.
Stellar evolution tracks without mass loss and convective core overshooting tend to have wider blue loops than model  with Cepheid mass loss and $\alpha_c=0.335$ for the same initial mass, where the width is defined by the change of effective temperature across the blue loop.  One might consider that the shape of the blue loops is an argument against Cepheid mass loss, but, the question of why a blue loop occurs at all is still unanswered.   \cite{Valle2009} found that the blue loop structure is affected by the assumed stellar mass, composition, and amount of convective core overshooting. We refer the reader to \cite{Neilson2011} for more discussion.

\begin{figure}[t]
\begin{center}
\includegraphics[width=0.7\textwidth]{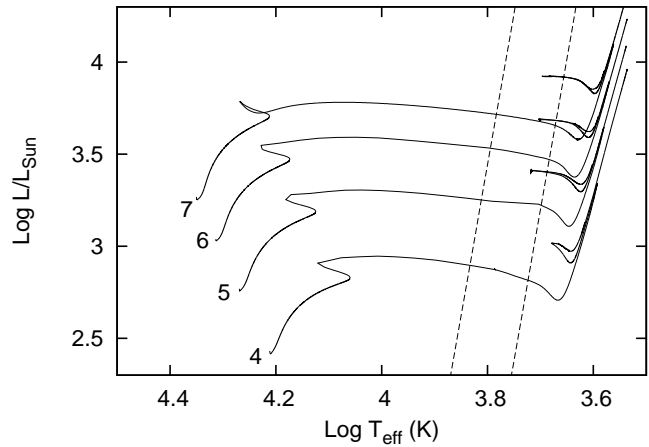}
\end{center}
\caption{Stellar evolution tracks for models assuming the standard LMC metallicity, $Z=0.008$.  The dashed lines denote the assumed boundaries of the Cepheid instability strip. }\label{f2}
\end{figure}
We repeat the calculations from \cite{Neilson2011} for stellar evolution models with LMC metallicity, $Z=0.008$.  In the \cite{Yoon2005} stellar evolution code, the helium abundance is correlated with the metallicity, therefore LMC stellar evolution models also assume a smaller helium abundance than Galactic models.  Evolution models are computed assuming $\alpha_c = 0.335$ and initial masses, $M=4,5,6,$ and $7~M_\odot$, with evolution tracks shown in Fig.~\ref{f2}, along with the estimated boundaries of the instability strip defined by \cite{Bono2000}.  These are the boundaries determined for Galactic Cepheids, however we demonstrated \citep{Neilson2011} that the total amount of mass lost during the blue loop evolution is largely insensitive to assumed location of the red edge of the instability strip.

The mass-loss theory developed by \cite{Neilson2008}, based on the \cite{Castor1975} CAK radiative-driven wind theory, noted that Cepheid mass-loss rates explicitly depend on metallicity, $\dot{M} \propto (Z/0.02)^{1/2}$. Therefore, we would expect that LMC Cepheid mass-loss rates  to be about $0.63\times$ mass-loss rates for Galactic Cepheids, if all other properties are the same.   Assuming the evolutionary timescale of the Cepheid blue loop is the same for both Galactic and LMC metallicities then we would expect the Cepheid mass discrepancy for LMC Cepheids to be about $3-4\%$ smaller than that for Galactic Cepheids. Furthermore, since the LMC models have smaller helium abundances then the luminosity during blue loop evolution is also smaller.  Therefore, the expected  difference between Galactic and LMC mass discrepancies would be $> 3-4\%$, while the observed difference from \cite{Keller2006} is about $-3\%$.
\begin{figure}[t]
\begin{center}
\includegraphics[width=0.8\textwidth]{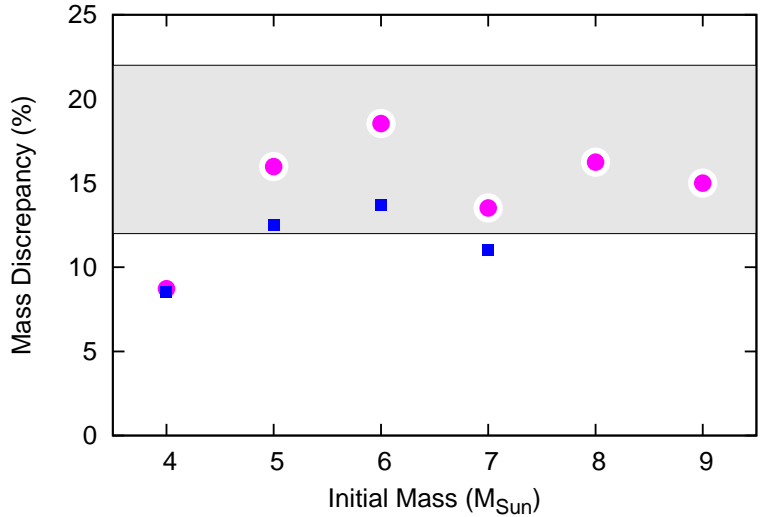}
\end{center}
\caption{Predicted Cepheid mass discrepancy for Galactic (magenta circles) and LMC (blue squares) as function of initial stellar mass.  The shaded region represents the measured Cepheid mass discrepancy of $17\pm5\%$ for Galactic Cepheids \citep{Keller2008}.}\label{f3}
\end{figure}

In Fig.~\ref{f3}, we show the predicted mass discrepancy for stellar evolution models with Galactic and LMC metallicities for the assumed stellar masses in this work.  We also plot the measured mass discrepancy for Galactic Cepheids from \cite{Keller2008}, $17\pm5\%$, as a shaded box to account for the uncertainties.  The mass-loss rates for LMC Cepheids tend to be smaller than for Galactic Cepheids leading to a smaller Cepheid mass discrepancy with decreasing metallicity. The difference between predicted Galactic and LMC mass discrepancy ranges from about zero at $4~M_\odot$ to $7\%$ at $6~M_\odot$.  However, the difference is not a function of initial mass, at $7~M_\odot$ the difference is about $2\%$.

The results presented in Fig.~\ref{f3} are not consistent with the metallicity dependence of the mass discrepancy determined by \cite{Keller2006}.  However, the predicted mass discrepancy as a function of metallicity is different than naively expected, suggesting that differences in stellar properties due to assumed metallicity are acting to predict larger mass-loss rates at smaller metallicities.  Another issue is the use of the period-pulsation amplitude relations calibrated using Galactic Cepheids;  \cite{Bono2000} found that pulsation models of LMC Cepheids have larger amplitudes than similar mass Galactic Cepheids. From \cite{Neilson2008}, an increase of the brightness amplitude by a factor of 2 leads to an increase of mass loss by about $5\%$, which leads to a significantly increased predicted mass discrepancy at LMC metallicity. This does not even include the effect of a larger radius amplitude on the predicted mass-loss rates. This behavior is consistent with the results of \cite{Neilson2009}, where mass-loss rates are computed for \cite{Bono2000} theoretical models and found larger mass-loss rates for LMC composition relative to Galactic metallicity.

\section{Summary}
In this work, we computed stellar evolution models with LMC metallicity using the \cite{Yoon2005} stellar evolution code that includes the \cite{Neilson2008} prescription for Cepheid mass loss.  From these models, we tested how much of the measured Cepheid mass discrepancy \citep{Keller2006} can be accounted for by assuming moderate convective core overshooting and pulsation-driven mass loss. We also compared these predictions with mass discrepancy predictions for Galactic Cepheids from \cite{Neilson2011} to see if we agree with the metallicity dependence of the mass discrepancy found by \cite{Keller2006}.

The results are not precise enough to demonstrate that pulsation-driven mass loss plus moderate convective core overshooting, $\alpha_c =0.335$, is the solution to the metallicity dependence of the Cepheid mass discrepancy.  However, by assuming the pulsation amplitudes are the same for Galactic and LMC metallicities, we have underestimated both the pulsation amplitudes and mass-loss rates. Thus, we underestimated the contribution of Cepheid mass loss to the LMC Cepheid mass discrepancy, hence, we cannot rule out the possibility that pulsation-driven mass loss is consistent with the results of \cite{Keller2006}.  This is encouraging since one would naively expect significantly smaller mass-loss rates for smaller metallicities, which we do not find.  Pulsation-driven mass loss plus convective core overshooting provides a plausiable solution to the Cepheid mass discrepancy, but more precise studies are required to verify this hypothesis.

\acknowledgements HRN is grateful for funding from the Alexander von Humboldt Foundation.

\bibliographystyle{asp2010}
\bibliography{lmc}

\end{document}